\documentclass[
 reprint,
 amsmath,amssymb,
 aps,superscriptaddress,
]{revtex4-2}

\usepackage{graphicx}
\usepackage{amssymb, amsmath, amsthm}
\usepackage{booktabs}
\usepackage[usenames]{color}
\usepackage{bm}
\usepackage{multirow}
\usepackage{dcolumn}
\usepackage{hyperref}
\usepackage{enumerate}
\usepackage{siunitx}
\usepackage{bbding}
\hypersetup{
    colorlinks=true,
    linkcolor=blue,
    filecolor=gray,
    urlcolor=blue,
    citecolor=blue,
}

\begin{document}
\title{Universal Hamiltonian learning with rotational and basis equivariance}

\newcommand{\thuphy}{State Key Laboratory of Low Dimensional Quantum Physics and Department of Physics, Tsinghua University, Beijing, 100084, China}
\newcommand{\thuias}{Chen-Ning Yang Institute for Advanced Study, Tsinghua University, Beijing 100084, China}
\newcommand{\fscqi}{Frontier Science Center for Quantum Information, Beijing 100084, China}
\newcommand{\srcComment}[1]{}

\affiliation{\thuphy}
\affiliation{\thuias}
\affiliation{\fscqi}

\author{Yuxiang~\surname{Wang}}
\thanks{These authors contributed equally to this work.}
\affiliation{\thuphy}
\affiliation{\thuias}

\author{He~\surname{Li}}
\thanks{These authors contributed equally to this work.}
\affiliation{\thuphy}
\affiliation{\thuias}

\author{Zechen~\surname{Tang}}
\thanks{These authors contributed equally to this work.}
\affiliation{\thuphy}

\author{Yang~\surname{Li}}
\affiliation{\thuphy}

\author{Honggeng~\surname{Tao}}
\affiliation{\thuphy}

\author{Yanzhen~\surname{Wang}}
\affiliation{\thuphy}

\author{Zilong~\surname{Yuan}}
\affiliation{\thuphy}

\author{Zezhou~\surname{Chen}}
\affiliation{\thuphy}

\author{Wenhui~\surname{Duan}}
\email{duanw@tsinghua.edu.cn}
\affiliation{\thuphy}
\affiliation{\thuias}
\affiliation{\fscqi}

\author{Yong~\surname{Xu}}
\email{yongxu@mail.tsinghua.edu.cn}
\affiliation{\thuphy}
\affiliation{\fscqi}

\begin{abstract}
Deep learning has the potential to transform computational materials research by making electronic-structure calculations more efficient and scalable. The Hamiltonian matrix in the localized atomic-orbital basis is a promising target for deep-learning electronic structure, yet enforcing rotational equivariance can be computationally costly, and the matrix depends on the choice of basis set.
Here we introduce DeepH-2, a physics-guided neural-network framework that reduces the computational cost of enforcing rotational equivariance through a local-coordinate formulation and incorporates basis equivariance by mapping a shared latent Hamiltonian to the target atomic-orbital basis.
DeepH-2 achieves sub-meV test-set error on a diverse materials dataset and demonstrates transferability across material dimensions.
These results highlight the value of encoding physics priors into deep-learning electronic structure models, and pave the way for developing the electronic-structure foundation models.
\end{abstract}

\maketitle

Density functional theory (DFT) is a cornerstone of computational materials science, but its computational cost limits its application to large and complex material systems. Machine learning offers a route to overcome this bottleneck by learning from first-principles calculations and transferring the resulting models to structures that are too large for routine DFT calculations. Such transferability has been demonstrated for both atomic-structure models, including deep-learning force fields~\cite{Lorenz2004, Behler2007, Brockherde2017, Justin2017, Zhang2018, Schutt2018, Xie2018, Batzner2022, Musaelian2023}, and electronic-structure models that directly learn quantum-mechanical quantities, including molecular wavefunctions and electron densities~\cite{Schutt2019, Unke2021}, as well as Hamiltonians~\cite{hegde2017machine, li2018density, Li2022, Gong2023, Li2023, Li2022_2, Li2023_2, tang2023efficient, tang2025deep, qian2026equivariant}. These advances exploit the nearsightedness principle~\cite{Kohn2005}, enabling local models trained on relatively small systems to generalize to larger structures. However, transferability across system sizes does not by itself guarantee universality across materials. In particular, developing a universal electronic-structure model requires not only generalization across diverse chemical environments and geometries, but also the ability to accommodate different representations of the underlying quantum operators. Addressing this representation dependence is essential for extending machine-learning electronic structure beyond models tied to particular material classes or basis choices.

Deep-learning electronic-structure models typically learn quantum operators, such as the DFT Hamiltonian, which is commonly represented in a basis of localized numerical atomic orbitals (NAOs). This representation provides a physically meaningful and computationally convenient description of electronic structure, but also introduces two distinct yet interconnected challenges for universal Hamiltonian learning. The first concerns angular structure. The angular components of NAOs are described by spherical harmonics, whose transformation properties under spatial rotations are determined by angular momentum. Accordingly, the mapping from atomic structures to Hamiltonian matrices must respect the rotational symmetry of the underlying electronic structure, formulated as equivariance under the Euclidean group E(3). Existing approaches address this requirement through symmetry-informed invariant architectures such as deep-learning DFT Hamiltonian (DeepH)~\cite{Li2022} and explicitly E(3)-equivariant networks such as DeepH-E3~\cite{Gong2023}. Although explicit equivariance provides a systematic treatment of spatial transformations, its computational cost can become substantial when multiple high-angular-momentum channels are involved. Local-coordinate formulations offer a promising route to represent angular information more efficiently~\cite{passaro2023reducing, liao2023equiformerv2}. The second challenge arises from the radial structure of NAOs. Unlike their angular components, which follow a well-defined spherical-harmonic structure, radial functions can vary with the choice of atomic-orbital basis. Consequently, the same Hamiltonian operator can have different matrix representations in different bases, with their relationship governed by the corresponding basis transformations. Geometric equivariance alone does not resolve this representation dependence, limiting the transferability of models across different atomic bases. Incorporating the transformation properties of Hamiltonian representations under basis changes therefore provides a physical principle for constructing more universal electronic-structure models.

Here we introduce DeepH-2, a framework designed to address both challenges in universal DFT Hamiltonian learning. Building upon previous DeepH developments~\cite{Li2022,Gong2023}, we first introduce the equivariant local-coordinate transformer (ELCT), which performs equivariant message passing in bond-dependent local coordinate systems, enabling efficient treatment of angular information while reducing the computational overhead associated with full SO(3)-equivariant operations. We further introduce the equivariant basis transformation layer (EBTL), which constructs a latent Hamiltonian representation in a common latent atomic-basis space and maps it to element-specific atomic bases through equivariant basis transformations. This design enables feature sharing across diverse atomic-basis representations while preserving the relevant transformation properties of Hamiltonian matrices. Together, these components allow DeepH-2 to combine efficient geometric equivariance with physically informed basis adaptation. DeepH-2 achieves higher accuracy and greater computational efficiency than previous Hamiltonian-learning models. Moreover, it exhibits zero-shot transferability to previously unseen systems, including carbon nanotubes and a twisted BN-graphene heterostructure, without additional training or fine-tuning. These results establish a promising route toward universal Hamiltonian learning across chemically and structurally diverse materials by addressing both the geometric symmetry and basis dependence of electronic-structure representations.

\begin{figure}
\includegraphics[width=1.0\linewidth]{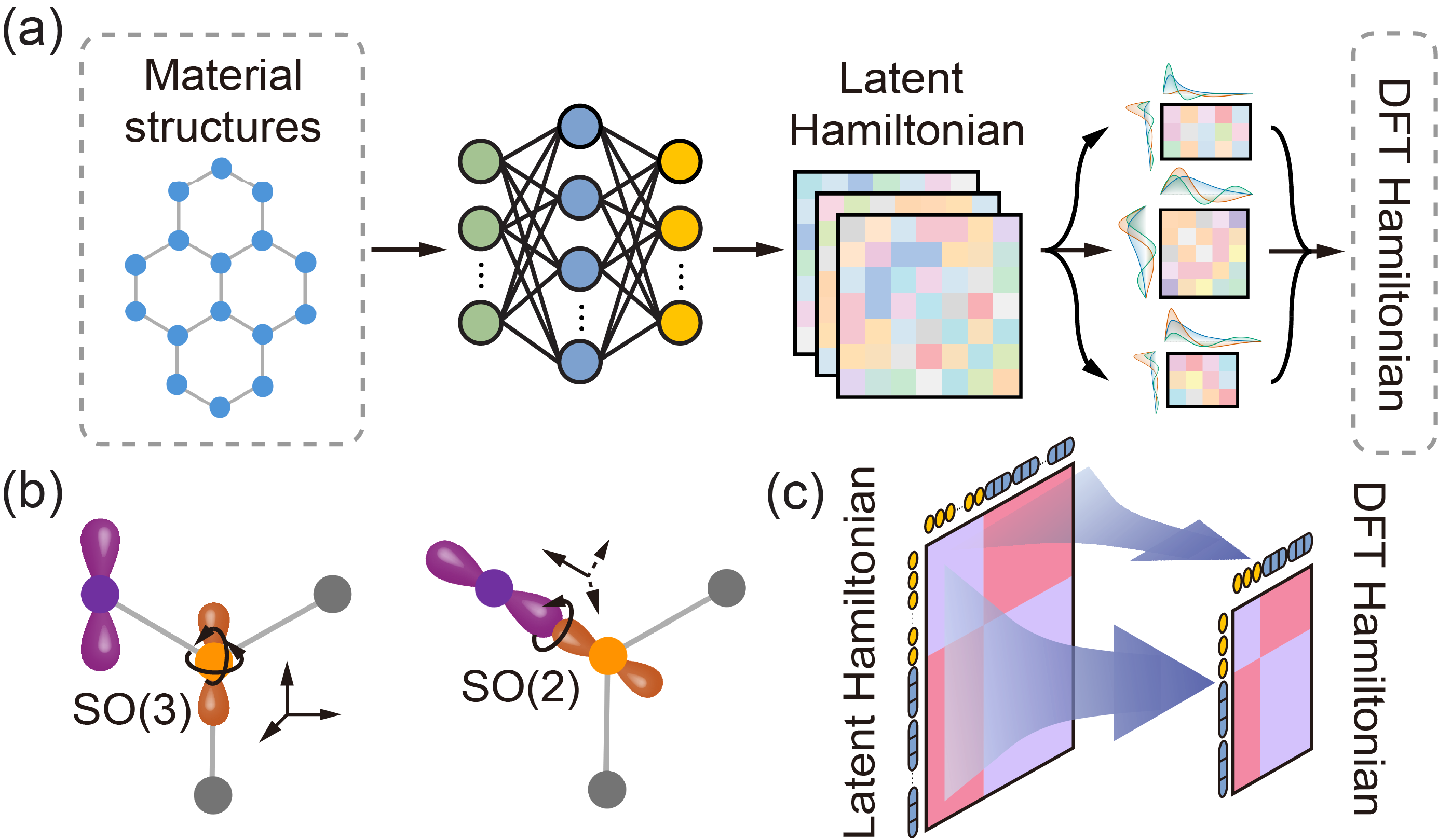}
\caption{\label{fig:architecture} 
Workflow and physics priors in DeepH-2. (a) Given a material structure, a neural network predicts a latent Hamiltonian, which is transformed into the density functional theory (DFT) Hamiltonian in the atomic-orbital basis. (b) The equivariant local coordinate framework uses bond-aligned frames to simplify global SO(3) equivariance to local SO(2) equivariance about each bond axis. (c) To account for the dependence of the Hamiltonian matrix on the choice of atomic-orbital basis, DeepH-2 first predicts a latent Hamiltonian with no explicit basis choice, then uses the equivariant basis transformation layer to obtain the DFT Hamiltonian in the atomic basis.
}
\end{figure}

\begin{figure}
\includegraphics[width=1.0\linewidth]{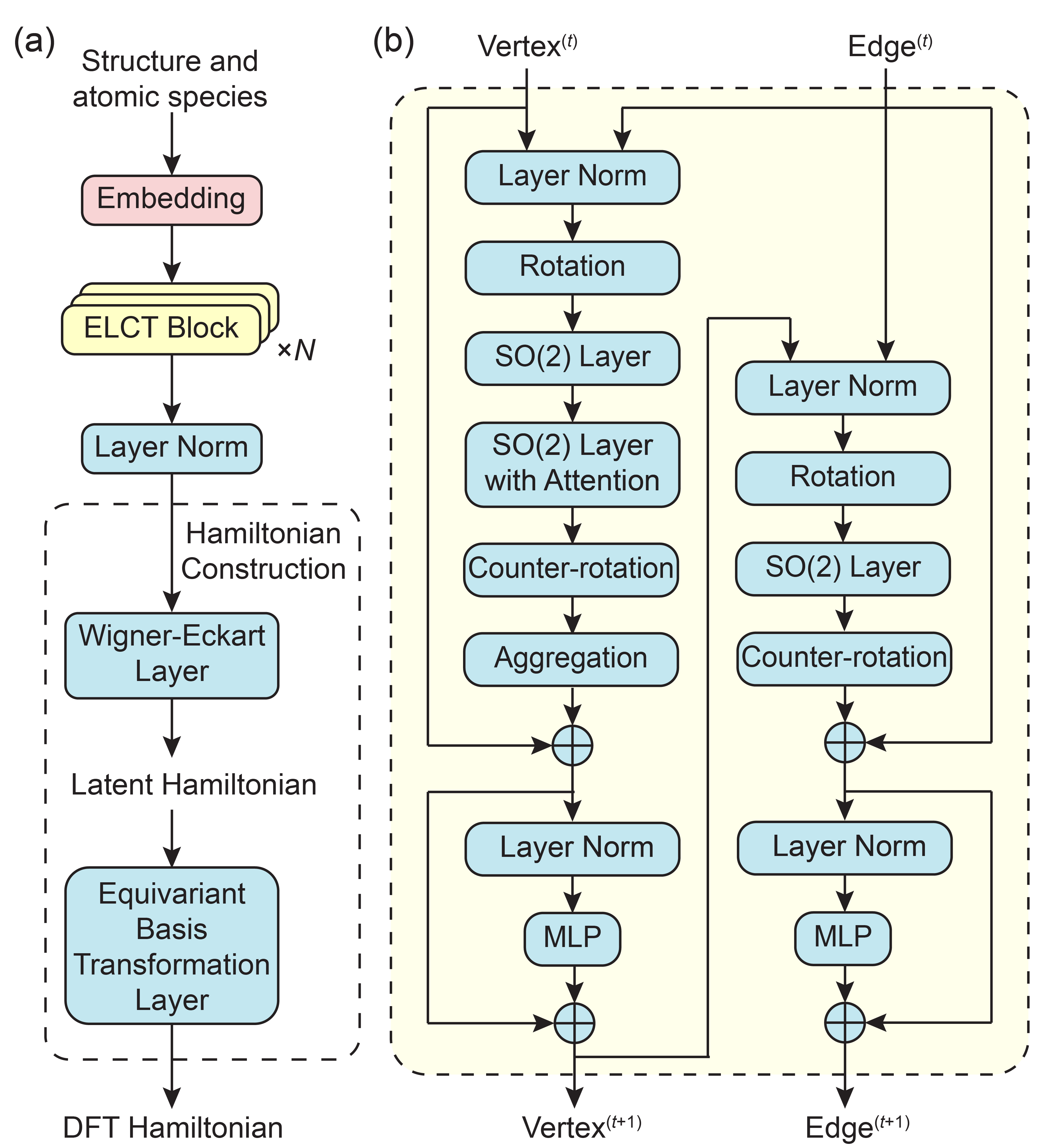}
\caption{\label{fig:arch1} Neural-network architecture of DeepH-2.
(a) Overall architecture of DeepH-2. The atomic structure is encoded as vertex and edge features, which are updated by a stack of equivariant local-coordinate transformer (ELCT) blocks. A Wigner-Eckart layer~\cite{Gong2023} constructs the latent Hamiltonian from the resulting features. The equivariant basis transformation layer then maps it to the density functional theory (DFT) Hamiltonian in the target atomic-orbital basis.
(b) Architecture of the $t$-th ELCT block. For the vertex update, features are rotated into bond-aligned frames and processed by SO(2)-equivariant layers, with attention applied in the local frames. The resulting messages are rotated back to the global frame before aggregation. The edge update uses the updated vertex features and also performs SO(2)-equivariant processing in local frames before returning to the global frame.}
\end{figure}

An overview of the DeepH-2 architecture is shown in Fig.~\ref{fig:architecture}. DeepH-2 addresses two key challenges in neural-network learning of DFT Hamiltonians: efficiently capturing rotational and geometric information, and generalizing across diverse atomic-orbital bases. As illustrated in Fig.~\ref{fig:architecture}(a), the network takes atomic positions and species as input and constructs a graph in which neighboring atoms are connected within a cutoff distance. The structural information is then encoded into equivariant vertex and edge features, which are refined through a series of message-passing layers. The ELCT, illustrated in Fig.~\ref{fig:architecture}(b), processes features in local coordinate systems aligned with interatomic directions, enabling efficient modeling of angular dependencies through SO(2)-equivariant operations while preserving the required rotational symmetry. After successive ELCT blocks, the resulting features are mapped to a latent Hamiltonian representation. To accommodate different atomic-orbital bases, EBTL, shown in Fig.~\ref{fig:architecture}(c), transforms this latent Hamiltonian into the target element-specific bases using learned basis transformations applied to both orbital indices. By separating the shared latent representation from the element-dependent basis transformations, EBTL facilitates information sharing across different atomic bases while respecting their transformation properties. Together, ELCT and EBTL enable efficient and accurate Hamiltonian prediction across diverse material systems. We describe the two components in detail below.

The detailed network architecture of DeepH-2 is illustrated in Fig.~\ref{fig:arch1}. As shown in Fig.~\ref{fig:arch1}(a), the network takes the material structure $\{\mathcal{R}\}$, comprising atomic positions $\mathbf{R}_i$ and atomic species indices $Z_i$, as input and represents the system as a graph, where only atom pairs within a cutoff distance $R_{\mathrm{C}}$ are connected. The structural information is first encoded into equivariant vertex and edge features, which are then updated through $N$ stacked ELCT blocks. After a final layer normalization, a Wigner-Eckart layer constructs the latent Hamiltonian, which is subsequently mapped to the target atomic-orbital basis through EBTL.

As illustrated in Fig.~\ref{fig:arch1}(b), each ELCT block first updates the vertex features and then uses the updated vertex features to update the edge features. Inspired by EquiformerV2~\cite{liao2023equiformerv2}, ELCT employs bond-dependent local coordinate systems aligned with the interatomic directions $\hat{\mathbf{R}}_{ij}$. The features are first normalized and rotated into the local frame, where SO(2)-equivariant layers process their angular information. For vertex updates, an additional SO(2)-equivariant attention layer adaptively weights messages from neighboring edges, followed by a counter-rotation that transforms the features back to the global frame. The resulting messages are aggregated to update the vertex features. In contrast, edge features are updated through a similar sequence of local-coordinate operations but without attention-based aggregation. Both pathways incorporate residual connections, layer normalization, and multilayer perceptrons (MLPs) to refine the features while facilitating stable information propagation.

By replacing computationally expensive full SO(3)-equivariant operations with local SO(2)-equivariant transformations, ELCT reduces the computational scaling with the maximum angular momentum $L$ from $\mathcal{O}(L^6)$ to $\mathcal{O}(L^3)$. This reduction is particularly beneficial for Hamiltonian learning involving high-angular-momentum orbitals. For example, when $f$ orbitals are included, the maximum angular momentum reaches $L=6$ without spin-orbit coupling (SOC) and $L=7$ with SOC. The local-coordinate formulation provides an efficient representation of the angular and geometric degrees of freedom relevant to atomic-orbital interactions, while the locality of ELCT ensures size extensiveness and scalability to large systems.

To address the diverse atomic bases encountered in universal materials models, we introduce EBTL, as illustrated in Fig.~\ref{fig:architecture}(c).
Previous DFT Hamiltonian models typically associate Hamiltonian matrix elements with fixed orbital indices.
However, the Hamiltonian is a physical operator independent of the particular choice of basis, whereas its matrix representation depends on that choice. When the basis functions are linearly transformed, the Hamiltonian matrix must transform accordingly with respect to its two orbital indices. This basis equivariance is a fundamental consequence of quantum mechanics and imposes a natural constraint on how Hamiltonian matrix elements should be represented.
To explicitly incorporate basis equivariance, EBTL introduces a common latent atomic-basis space shared across elements.

The preceding neural network first predicts a latent Hamiltonian in this common latent basis as $H'^{\,ij}_{\alpha\beta}$, where $
\alpha=(c_\alpha,l_\alpha,m_\alpha),
\beta=(c_\beta,l_\beta,m_\beta),
$ with $c$, $l$, and $m$ denoting the channel index, angular momentum, and SO(3) representation component, respectively.
The element-specific Hamiltonian is then obtained by an equivariant transformation from the latent basis to the atomic bases of the two interacting atoms:
$$
H^{ij}_{\alpha\beta}
=
\sum_{c'_\alpha,c'_\beta}
W^{(Z_i,l_\alpha)}_{c_\alpha c'_\alpha}
H'^{\,ij}_{c'_\alpha l_\alpha m_\alpha,\,
c'_\beta l_\beta m_\beta}
W^{(Z_j,l_\beta)}_{c_\beta c'_\beta}.
$$
Here, $W^{(Z,l)}$ parameterizes the transformation between the common latent basis and the element-specific basis for atomic number $Z$ and angular momentum $l$. It mixes channels only within the same $l$ sector while preserving the corresponding $m$ component, and therefore maintains SO(3) equivariance. Since the transformation is applied to both orbital indices, the resulting Hamiltonian transformation takes a bilinear form, as required for a matrix representation under a change of basis.

By representing the Hamiltonian in a common latent basis and learning element-dependent basis transformations, EBTL decouples the shared physical representation from the particular choice of atomic basis. This allows corresponding orbital subspaces across different elements to be related through learned linear combinations rather than fixed orbital indices, enabling effective information sharing while respecting basis equivariance.

Overall, ELCT and EBTL improve Hamiltonian representation: ELCT efficiently captures rotational and angular information, while EBTL addresses the basis-channel degrees of freedom associated with diverse atomic orbitals.
Together, they enable accurate and scalable Hamiltonian prediction across diverse materials systems.

We first validate the effectiveness of ELCT on specialized benchmark datasets, using DeepH-2 without EBTL to isolate the contribution of ELCT. Since EBTL is designed to address the challenges posed by diverse atomic bases, its benefits are expected to be limited for these benchmarks, which involve relatively few types of atomic bases. This is consistent with the ablation results in Table~S3\srcComment{tab:table2}~\cite{supp}, where incorporating EBTL leads to only small changes in Hamiltonian MAE across monolayer and bilayer graphene and MoS$_2$. Without EBTL, DeepH-2 achieves substantially lower Hamiltonian mean absolute error (MAE) than both DeepH and DeepH-E3, reaching sub-$0.1$~meV Hamiltonian MAE across all four datasets (Table~S2\srcComment{tab:table1}~\cite{supp}). These results support the effectiveness of ELCT for accurate Hamiltonian prediction.

\begin{figure*}
\includegraphics[width=\textwidth]{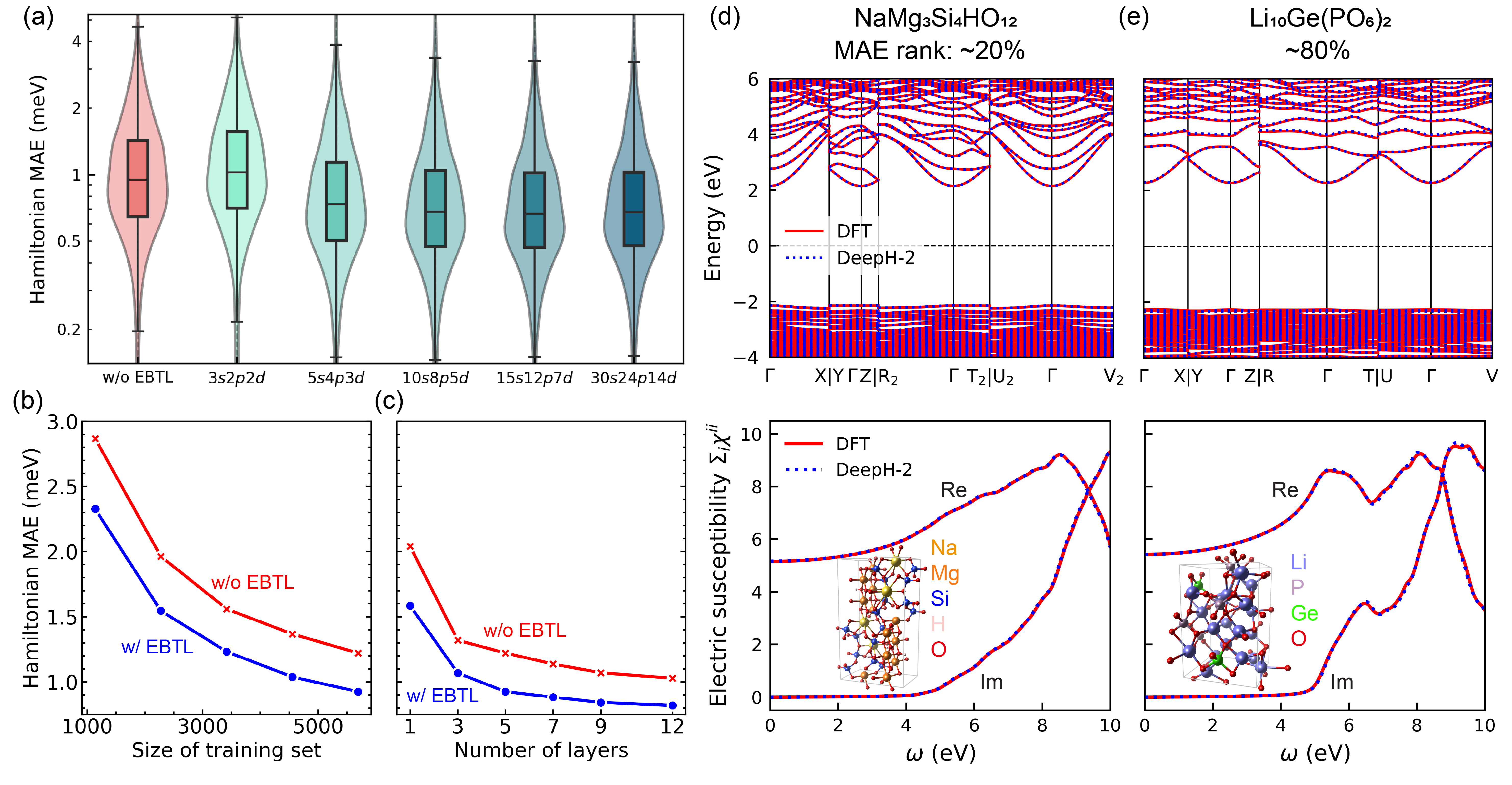}
\caption{\label{fig:mae_element} 
Effect of the equivariant basis transformation layer (EBTL) and test-set performance for the universal materials dataset.
(a) Violin plots of per-structure Hamiltonian mean absolute error (MAE) for different EBTL configurations. The overlaid box plots indicate the median, the 25th and 75th percentiles (box edges), and whiskers extending to the most extreme observations within $1.5\times$ the interquartile range; outliers are not shown. The MAE decreases as the latent space size grows and converges for larger bases. The default latent basis (15$s$12$p$7$d$) yields a mean MAE of 0.925~meV, compared with 1.221~meV without EBTL.
(b, c) Models with EBTL (blue) consistently achieve lower structure-averaged MAEs than models without EBTL (red) across training-set sizes and network depths.
(d, e) Density functional theory (DFT) and DeepH-2 band structures and electric susceptibilities $\chi(\omega)$ for NaMg$_3$Si$_4$HO$_{12}$ (mp-1221204) and Li$_{10}$Ge(PO$_6$)$_2$ (mp-632815), at approximately the 20th and 80th percentiles of the test-set MAE distribution, respectively. Insets show the corresponding
atomic structures.
}
\end{figure*}

We next evaluate the performance of DeepH-2 on the universal materials dataset~\cite{2024-deeph-umm} and systematically investigate the effectiveness of EBTL through an ablation study. Unlike previous DFT Hamiltonian prediction studies, which typically considered only a limited number of elements and atomic orbitals, universal materials datasets span a much broader range of chemical species, resulting in hundreds or even thousands of distinct atomic orbitals. Such diversity makes the Hamiltonian representation increasingly dependent on the choice of atomic basis, and therefore calls for an explicit treatment of basis equivariance to enable effective information sharing across different bases.
As shown in Fig.~\ref{fig:mae_element}(a), the Hamiltonian MAE depends on the size of EBTL. Models without EBTL or with a small EBTL exhibit substantially larger errors, whereas the MAE saturates when the EBTL is sufficiently large. These results demonstrate the importance of basis equivariance for accurate Hamiltonian prediction across a large number of atomic bases.
The benefit of EBTL remains robust across different training-set sizes, network depths, and training lengths (Figs.~\ref{fig:mae_element}(b,c) and S1\srcComment{figs1}(e)~\cite{supp}). In all three settings, models equipped with EBTL consistently achieve lower test-set MAEs than their counterparts without EBTL, demonstrating that the performance advantage of EBTL is robust to variations in both model architecture and training conditions.
Beyond the Hamiltonian-level accuracy, we further assess whether the predicted Hamiltonians enable high-precision calculations of physical properties. As shown in Figs.~\ref{fig:mae_element}(d,e) and S2\srcComment{figs62}(a,b)~\cite{supp}, DeepH-2 reproduces the DFT band structures and captures the overall frequency dependence of the electric susceptibility $\chi(\omega)$ for representative test-set materials. These results show that the low Hamiltonian MAE translates into accurate predictions of relevant electronic and response properties.
Overall, DeepH-2 provides accurate and robust Hamiltonian predictions across a diverse universal test set, establishing a basis for evaluating its transferability to unseen material classes.

\begin{figure}
\includegraphics[width=\linewidth]{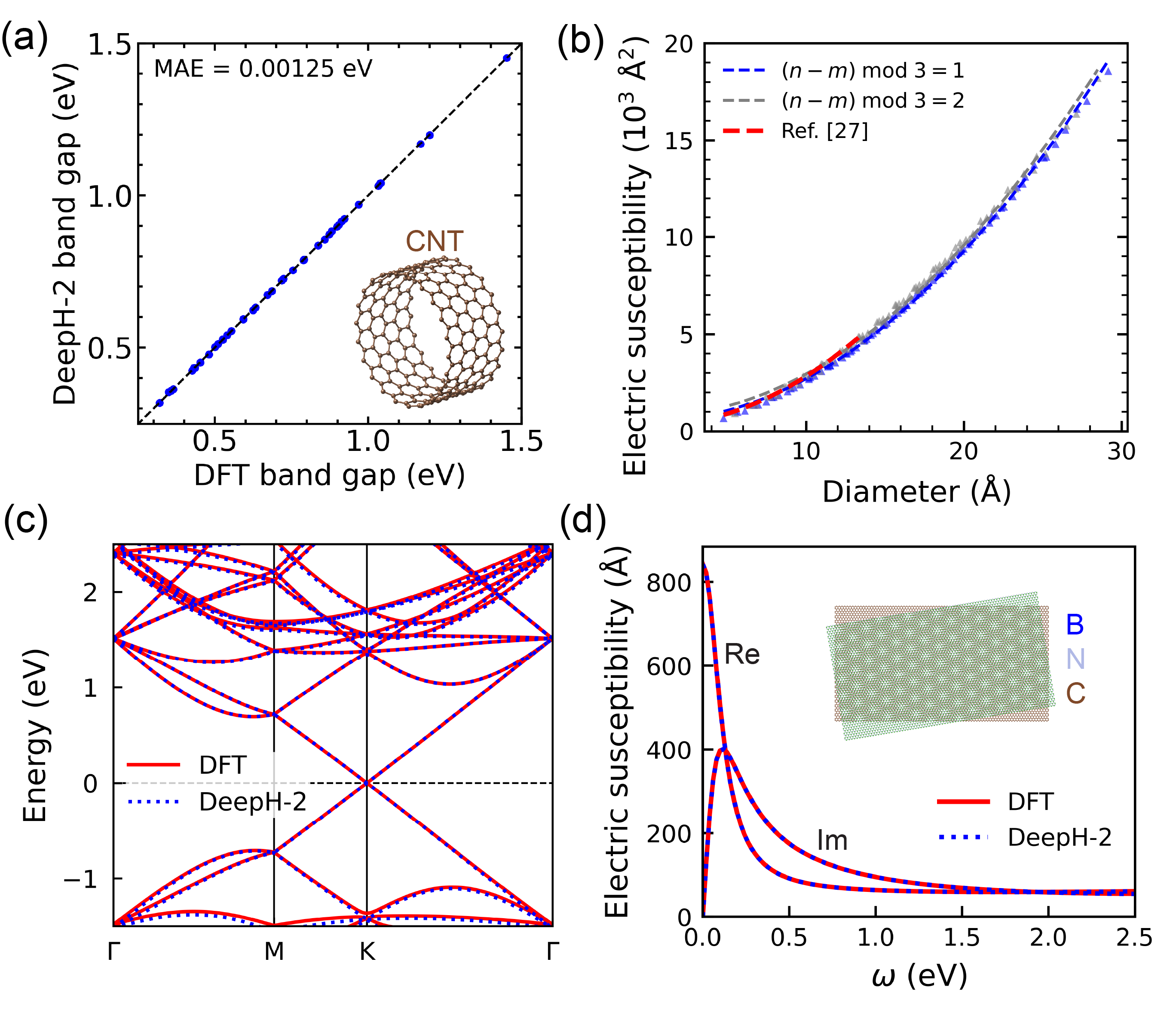}
\caption{\label{fig:transfer_ability} 
Zero-shot generalization of DeepH-2, trained predominantly on bulk materials, to low-dimensional materials.
(a) DeepH-2 and density functional theory (DFT) band gaps for 41 insulating carbon nanotubes (CNTs) with fewer than 495 atoms.
(b) Static electric susceptibility versus diameter for insulating CNTs. The red dashed line shows the relation fitted by Kozinsky and Marzari to DFT results for small-sized CNTs~\cite{Kozinsky2006}. Blue and gray dashed lines show quadratic fits for CNTs with $(n-m) \bmod 3 = 1$ and $2$, respectively.
(c, d) Comparison of DeepH-2 and DFT results for the $(4,3)$ twisted BN-graphene heterostructure. The band structure (c) and electric susceptibility (d) both show close agreement. The inset in (d) shows the atomic structure.
}
\end{figure}

To demonstrate transferability of DeepH-2, we evaluate its zero-shot generalization from predominantly bulk training data to carbon nanotubes (CNTs) and a twisted BN-graphene heterostructure, without any additional training or fine-tuning on low-dimensional systems. The results are summarized in Fig.~\ref{fig:transfer_ability}.

We first consider CNTs as a test of structural transferability across material dimensions. Unlike the predominantly bulk structures in the universal training dataset, CNTs exhibit strongly curved geometries as well as a wide range of diameters and system sizes. Compared with the original DeepH study~\cite{Li2022}, DeepH-2 achieves substantially lower Hamiltonian matrix MAEs for both armchair and zigzag CNTs, as shown in Fig.~S4\srcComment{figs72}~\cite{supp}. Importantly, its accuracy remains more robust as the CNT diameter decreases, despite the increasingly strong curvature of the structures. In contrast, the performance of DeepH deteriorates for highly curved CNTs, which can be attributed to the limited representation of strongly curved local environments in its graphene-based training dataset~\cite{Li2022}. Notably, the universal dataset used to train DeepH-2 contains no CNT structures and only a limited number of carbon allotropes, with the vast majority of its structures being bulk materials. The ability of DeepH-2 to accurately predict the Hamiltonians of CNTs therefore suggests that its learned representation captures transferable features of local chemical environments rather than relying on system-specific structural patterns.

We next examine the predictions of physical properties. As shown in Figs.~\ref{fig:transfer_ability}(a) and S5\srcComment{figs73}(b)~\cite{supp}, the physical properties predicted by DeepH-2 agree well with DFT calculations. In particular, Fig.~\ref{fig:transfer_ability}(b) shows that the zero-shot predictions reproduce the quadratic dependence of the static electric susceptibility on the CNT diameter reported by Kozinsky and Marzari~\cite{Kozinsky2006}. With the larger dataset considered here, we further identify a systematic difference between CNTs with $(n-m)\bmod 3=1$ and $2$, which is captured by the DeepH-2 predictions. For larger CNTs, where direct DFT calculations become computationally prohibitive, DeepH-2 predicts the systematic size-dependent evolution of the electric susceptibility for systems containing up to 5,548 atoms in Fig.~S6\srcComment{figs74}~\cite{supp}. These results demonstrate transferability from bulk to strongly curved low-dimensional systems, together with scalability to substantially larger system sizes.

The cross-dimensional transferability is also examined on the twisted van der Waals heterostructure. As shown in Figs.~\ref{fig:transfer_ability}(c,d), the DeepH-2 predictions reproduce both the electronic band structure and the electric susceptibility of the twisted BN-graphene heterostructure obtained from DFT, despite the absence of such heterostructures from the training data. Together with the CNT results, these findings demonstrate that the DeepH-2 universal model can transfer directly between substantially different material geometries and dimensionalities without system-specific retraining. The successful predictions across different geometries, dimensionalities, and system sizes highlight the robustness and generality of the universal model. This zero-shot capability enables the direct application of DeepH-2 to diverse materials beyond the training distribution, without constructing system-specific training datasets.

To summarize, we developed DeepH-2, a physics-guided deep-learning framework for DFT Hamiltonians that incorporates rotational and basis equivariance. Rotational equivariance is addressed by the ELCT framework, which preserves global SO(3) equivariance by performing more efficient SO(2)-equivariant message passing in bond-aligned frames. The EBTL architecture accounts for the dependence of Hamiltonian matrices on the choice of atomic basis by mapping a shared latent Hamiltonian to element-specific atomic-orbital bases. These two architectural advances significantly increase the predictive power of DeepH-2, which shows promise as a transferable and scalable framework for learning Hamiltonian representations beyond specific material classes. These developments provide a foundation for further advancing universal electronic-structure learning and extending its applicability to a broader range of materials systems.

\begin{acknowledgments}
This work was supported by the National Key Basic Research and Development Program of China (Grant No. 2024YFA1409100), the Basic Science Center Project of NSFC (Grant No. 52388201), the National Natural Science Foundation of China (Grant Nos. 12334003, 12421004, 124B2072, and 12361141826), the National Key Basic Research and Development Program of China (Grant No. 2023YFA1406400), Quantum Science and Technology---National Science and Technology Major Project (Grant No. 2023ZD0300500), Fundamental and Interdisciplinary Disciplines Breakthrough Plan of the Ministry of Education of China (Grant No. JYB2025XDXM408), the China Postdoctoral Science Foundation (grant no. 2025M773367), ShanghaiTech AI4S Initiative SHTAI4S202504, Beijing Key Laboratory of Quantum AI, and the Beijing Advanced Innovation Center for Future Chip (ICFC). Yang Li is funded by the Shuimu Tsinghua Scholar program. The calculations were performed at National Supercomputer Center in Tianjin using the Tianhe new generation supercomputer.
\end{acknowledgments}

%

\end{document}